\documentclass{aastex61}
\usepackage{amsmath}
\usepackage{graphicx}
\usepackage{comment}
\usepackage{color}
\usepackage{bm}

\shorttitle{MOA-2015-BLG-337}
\shortauthors{Miyazaki et al.}

\begin{document}

\title{MOA-2015-BLG-337: A PLANETARY SYSTEM WITH A LOW-MASS BROWN DWARF/PLANETARY BOUNDARY HOST, OR A BROWN DWARF BINARY}

\author[0000-0002-0786-7307]{S. Miyazaki$^{\dag}$}
\affil{Department of Earth and Space Science, Graduate School of Science, Osaka University, 1-1 Machikaneyama, Toyonaka, Osaka 560-0043, Japan}
\affil{MOA collaboration}
\author{T. Sumi}
\affil{Department of Earth and Space Science, Graduate School of Science, Osaka University, 1-1 Machikaneyama, Toyonaka, Osaka 560-0043, Japan}
\affil{MOA collaboration}
\author{D. P. Bennett}
\affil{Department of Physics, University of Notre Dame, Notre Dame, IN 46556, USA} 
\affil{Laboratory for Exoplanets and Stellar Astrophysics, NASA/Goddard Space Flight Center, Greenbelt, MD 20771, USA}
\affil{MOA collaboration}
\author{A. Gould}
\affil{Korea Astronomy and Space Science Institute, Daejon 34055, Korea}
\affil{Department of Astronomy Ohio State University, 140 W. 18th Ave., Columbus, OH 43210, USA}
\affil{Max-Planck-Institute for Astronomy, K\"{o}nigstuhl 17, 69117 Heidelberg, Germany}
\affil{KMTNet collaboration}
\author{A. Udalski}
\affil{Warsaw University Observatory, Al.~Ujazdowskie~4, 00-478~Warszawa, Poland}
\affil{OGLE collaboration}
\author{I. A. Bond}
\affil{Institute of Information and Mathematical Sciences, Massey University, Private Bag 102-904, North Shore Mail Centre, Auckland, New Zealand} 
\affil{MOA collaboration}
\author{N. Koshimoto}
\affil{Department of Earth and Space Science, Graduate School of Science, Osaka University, 1-1 Machikaneyama, Toyonaka, Osaka 560-0043, Japan}
\affil{MOA collaboration}
\author{M. Nagakane}
\affil{Department of Earth and Space Science, Graduate School of Science, Osaka University, 1-1 Machikaneyama, Toyonaka, Osaka 560-0043, Japan}
\affil{MOA collaboration}
\author{N. Rattenbury}
\affil{Department of Physics, University of Auckland, Private Bag 92019, Auckland, New Zealand}
\affil{MOA collaboration}
\nocollaboration

\author{F. Abe}
\affil{Institute for Space-Earth Environmental Research, Nagoya University, Nagoya 464-8601, Japan} 
\author{A. Bhattacharya}
\affil{Department of Physics, University of Notre Dame, Notre Dame, IN 46556, USA} 
\affil{Laboratory for Exoplanets and Stellar Astrophysics, NASA/Goddard Space Flight Center, Greenbelt, MD 20771, USA} 
\author{R. Barry}
\affil{Laboratory for Exoplanets and Stellar Astrophysics, NASA/Goddard Space Flight Center, Greenbelt, MD 20771, USA}
\author{M. Donachie}
\affil{Department of Physics, University of Auckland, Private Bag 92019, Auckland, New Zealand}
\author{A. Fukui}
\affil{Okayama Astrophysical Observatory, National Astronomical Observatory, 3037-5 Honjo, Kamogata, Asakuchi, Okayama 719-0232, Japan}
\author{Y. Hirao}
\affil{Department of Earth and Space Science, Graduate School of Science, Osaka University, 1-1 Machikaneyama, Toyonaka, Osaka 560-0043, Japan}
\author{Y. Itow}
\affil{Institute for Space-Earth Environmental Research, Nagoya University, Nagoya, 464-8601, Japan}
\author{K. Kawasaki}
\affil{Department of Earth and Space Science, Graduate School of Science, Osaka University, 1-1 Machikaneyama, Toyonaka, Osaka 560-0043, Japan}
\author{M. C. A. Li}
\affil{Department of Physics, University of Auckland, Private Bag 92019, Auckland, New Zealand}
\author{C. H. Ling}
\affil{Institute of Information and Mathematical Sciences, Massey University, Private Bag 102-904, North Shore Mail Centre, Auckland, New Zealand}
\author{Y. Matsubara}
\affil{Institute for Space-Earth Environmental Research, Nagoya University, Nagoya, 464-8601, Japan}
\author{T. Matsuo}
\affil{Department of Earth and Space Science, Graduate School of Science, Osaka University, 1-1 Machikaneyama, Toyonaka, Osaka 560-0043, Japan}
\author{Y. Muraki}
\affil{Institute for Space-Earth Environmental Research, Nagoya University, Nagoya, 464-8601, Japan}
\author{K. Ohnishi}
\affil{Nagano National College of Technology, Nagano 381-8550, Japan}
\author{C. Ranc}
\affil{Laboratory for Exoplanets and Stellar Astrophysics, NASA/Goddard Space Flight Center, Greenbelt, MD 20771, USA}
\author{T. Saito}
\affil{Tokyo Metropolitan College of Industrial Technology, Tokyo 116-8523, Japan}
\author{A. Sharan}
\affil{Department of Physics, University of Auckland, Private Bag 92019, Auckland, New Zealand}
\author{H. Shibai}
\affil{Department of Earth and Space Science, Graduate School of Science, Osaka University, 1-1 Machikaneyama, Toyonaka, Osaka 560-0043, Japan}
\author{H. Suematsu}
\affil{Department of Earth and Space Science, Graduate School of Science, Osaka University, 1-1 Machikaneyama, Toyonaka, Osaka 560-0043, Japan} 
\author{D. Suzuki}
\affil{Institute of Space and Astronautical Science, Japan Aerospace Exploration Agency, 3-1-1 Yoshinodai, Chuo, Sagamihara, Kanagawa 252-5210, Japan}
\author{D.J. Sullivan}
\affil{School of Chemical and Physical Sciences, Victoria University, Wellington, New Zealand}
\author{P. J. Tristram}
\affil{University of Canterbury Mt. John Observatory, P.O. Box 56, Lake Tekapo 8770, New Zealand}
\author{T. Yamada}
\affil{Department of Earth and Space Science, Graduate School of Science, Osaka University, 1-1 Machikaneyama, Toyonaka, Osaka 560-0043, Japan}
\author{A. Yonehara}
\affil{Department of Physics, Faculty of Science, Kyoto Sangyo University, Kyoto 603-8555, Japan}
\collaboration{(MOA collaboration)}

\author{S. Koz\L owski}
\affil{Warsaw University Observatory, Al.~Ujazdowskie~4, 00-478~Warszawa, Poland}
\author{P. Mr\'oz}
\affil{Warsaw University Observatory, Al.~Ujazdowskie~4, 00-478~Warszawa, Poland}
\author{M. Pawlak}
\affil{Warsaw University Observatory, Al.~Ujazdowskie~4, 00-478~Warszawa, Poland}
\author{R. Poleski}
\affil{Department of Astronomy, Ohio State University, 140 W. 18th Ave., Columbus, OH  43210, USA}
\author{P. Pietrukowicz}
\affil{Warsaw University Observatory, Al.~Ujazdowskie~4, 00-478~Warszawa, Poland}
\author{J. Skowron}
\affil{Warsaw University Observatory, Al.~Ujazdowskie~4, 00-478~Warszawa, Poland}
\author{I. Soszy{\'n}ski}
\affil{Warsaw University Observatory, Al.~Ujazdowskie~4, 00-478~Warszawa, Poland}
\author{M. K. Szyma\'nski}
\affil{Warsaw University Observatory, Al.~Ujazdowskie~4, 00-478~Warszawa, Poland}
\author{K. Ulaczyk}
\affil{Warsaw University Observatory, Al.~Ujazdowskie~4, 00-478~Warszawa, Poland}
\collaboration{(OGLE collaboration)}

\author{M. D. Albrow}
\affil{University of Canterbury, Department of Physics and Astronomy, Private Bag 4800, Christchurch 8020, New Zealand}
\author{S.-J. Chung}
\affil{Korea Astronomy and Space Science Institute, Daejon 34055, Korea}
\affil{Korea University of Science and Technology, 217 Gajeong-ro, Yuseong-gu, Daejeon, 34113, Korea}
\author{C. Han}
\affil{Department of Physics, Chungbuk National University, Cheongju 28644, Republic of Korea}
\author{Y. K. Jung}
\affil{Korea Astronomy and Space Science Institute, Daejon 34055, Korea}
\author{K.-H. Hwang}
\affil{Korea Astronomy and Space Science Institute, Daejon 34055, Korea}
\author{Y.-H. Ryu}
\affil{Korea Astronomy and Space Science Institute, Daejon 34055, Republic of Korea}
\author{I.-G.Shin}
\affil{Harvard-Smithsonian Center for Astrophysics, 60 Garden St., Cambridge, MA 02138, USA}
\author{Y. Shvartzvald}
\affil{Jet Propulsion Laboratory, California Institute of Technology, 4800 Oak Grove Drive, Pasadena, CA 91109, USA}
\affil{NASA Postdoctoral Program Fellow}
\author{J. C. Yee}
\affil{Harvard-Smithsonian Center for Astrophysics, 60 Garden St., Cambridge, MA 02138, USA}
\author{W. Zang}
\affil{Physics Department and Tsinghua Centre for Astrophysics, Tsinghua University, Beijing 100084, China}
\affil{Department of Physics, Zhejiang University, Hangzhou, 310058, China}
\author{W. Zhu}
\affil{Canadian Institute for Theoretical Astrophysics, University of Toronto, Toronto, ON M5S 3H8, Canada}
\author{S.-M. Cha}
\affil{Korea Astronomy and Space Science Institute, Daejon 34055, Korea}
\affil{School of Space Research, Kyung Hee University, Yongin, Kyeonggi 17104, Korea}
\author{D.-J. Kim}
\affil{Korea Astronomy and Space Science Institute, Daejon 34055, Korea}
\author{H.-W. Kim}
\affil{Korea Astronomy and Space Science Institute, Daejon 34055, Korea}
\author{S.-L. Kim}
\affil{Korea Astronomy and Space Science Institute, Daejon 34055, Korea}
\affil{Korea University of Science and Technology, 217 Gajeong-ro, Yuseong-gu, Daejeon, 34113, Korea}
\author{C.-U. Lee}
\affil{Korea Astronomy and Space Science Institute, Daejon 34055, Korea}
\affil{Korea University of Science and Technology, 217 Gajeong-ro, Yuseong-gu, Daejeon, 34113, Korea}
\author{D.-J. Lee}
\affil{Korea Astronomy and Space Science Institute, Daejon 34055, Korea}
\author{Y. Lee}
\affil{Korea Astronomy and Space Science Institute, Daejon 34055, Korea}
\affil{School of Space Research, Kyung Hee University, Yongin, Kyeonggi 17104, Korea}
\author{B.-G. Park}
\affil{Korea Astronomy and Space Science Institute, Daejon 34055, Korea}
\affil{Korea University of Science and Technology, 217 Gajeong-ro, Yuseong-gu, Daejeon, 34113, Korea}
\author{R. W. Pogge}
\affil{Department of Astronomy Ohio State University, 140 W. 18th Ave., Columbus, OH 43210, USA}
\collaboration{(KMTNet collaboration)}

\begin{abstract}
\normalsize{
We report the discovery and the analysis of the short timescale binary-lens microlensing event, MOA-2015-BLG-337.
The lens system could be a planetary system with a very low mass host, around the brown dwarf/planetary mass boundary, or a brown dwarf binary. 
We found two competing models that explain the observed light curves with companion/host mass ratios of $q\sim0.01$ and $\sim0.17$, respectively.
A significant finite source effect in the best-fit planetary model ($q\sim0.01$) reveals a small angular Einstein radius of $\theta_{\rm E}\simeq0.03$ mas which favors a low mass lens.  
We obtain the posterior probability distribution of the lens properties from a Bayesian analysis.
The results for the planetary models strongly depend on a power-law index in planetary mass regime, $\alpha_{\rm pl}$, in the assumed mass function.
In summary, there are two solutions of the lens system: (1) a brown dwarf/planetary mass boundary object orbited by a super-Neptune (the planetary model with $\alpha_{\rm pl}=0.49$) and (2) a brown dwarf binary (the binary model).
If the planetary models are correct, this system can be one of a new class of planetary system, having a low host mass and also a planetary mass ratio ($q <0.03$) between the companion and its host.  
The discovery of the event is important for the study of planetary formation in very low mass objects.
In addition, it is important to consider all viable solutions in these kinds of ambiguous events in order for the future comprehensive statistical analyses of planetary/binary microlensing events.
}
\end{abstract}

\section{Introduction} \label{sec:intro}
\normalsize{
Exoplanets and brown dwarfs (BDs) are usually inferred from indirect methods owing to their intrinsic faintness.
Each method has its own unique sensitivity range and can probe different regions of the planet parameter space. 
Gravitational microlensing has an unequaled sensitivity to companions with masses ranging from stellar mass to even Earth mass planets orbiting beyond the snow line (\citealp{1991ApJ...374L..37M}; \citealp{1992ApJ...396..104G}).
According to the core accretion theory (\citealp{1996Icar..124...62P}; \citealp{2002ApJ...581..666K}), massive Jovian planets could be efficiently formed outside the snow line.
The radial velocity (RV) (\citealp{2006ApJ...646..505B}; \citealp{1995Natur.378..355M}) and transit methods (\citealp{2013ApJS..204...24B}; \citealp{2011ApJ...736...19B}) are mostly sensitive to planets that are relatively more massive or orbit closer to their host stars. 
\vspace{0.2in}

Searching for companions to BDs is very crucial in order for understanding of the formation mechanism of BDs.
Though still a matter of debate, most theories suppose that BDs were formed by the direct collapse of molecular clouds on a much smaller scale than stars, and promoted by turbulent fragmentation (\citealp{2012ARA&A..50...65L}).
Some young BDs are observed with excess emission from the surrounding raw material disks (\citealp{2005Sci...310..834A}; \citealp{2005ApJ...635L..93L}; \citealp{2012ApJ...761L..20R}).
The recent direct imaging survey to search for BDs and very low mass stars (VLMS)  found a strong preference to equal mass binary systems if they have companions (\citealp{2010ApJ...722..311L}; \citealp{2006ApJS..166..585B}).
\citet{2006ApJS..166..585B} found a very steep index $\gamma = 4.2 \pm 1.0$ for the power law distribution $\propto q^{\gamma}$, which is derived by fitting the observed mass ratio distribution.
Though there are still large uncertainties because the masses of the objects detected by direct imaging were estimated based on assumed ages, lower mass ratio (i.e., $q\le0.3$) binaries would be very uncommon among field VLMS and BDs. 
However, there have been several discoveries of binary BDs (\citealp{2017ApJ...843...59H}) as well as a BD orbiting an M dwarf (\citealp{2016ApJ...822...75H}) with small mass ratios ($q<0.3$) by microlensing.
In addition, \citet{2013ApJ...778...38H} found a system that consists of a several Jupiter-mass planet orbiting a BD and suggested that the planetary mass companion might have been formed in a  proto-planetary disk surrounding the BD.
Microlensing can play a meaningful role for understanding the formation of VLMS and BDs by detecting companions of low-mass hosts.
\vspace{0.2in}

Gravitational microlensing has been used to discover several dozen exoplanets around M-dwarf stars, revealing that massive Jovian planets beyond the snow line are much more uncommon compared to low-mass ones (\citealp{2010ApJ...720.1073G}; \citealp{2010ApJ...710.1641S}; \citealp{2012Natur.481..167C}; \citealp{2016MNRAS.457.4089S}).
Particularly, \citet{2016ApJ...833..145S} and \citet{2018arXiv180202582U} found a possible break and peak around $q\sim2\times10^{-4}$ in the companion-host mass ratio function.
They suggested that ice giant mass planets are the most abundant ones outside the snow line.
This is qualitatively consistent with predictions from core accretion theory.
However, in their results, the estimated abundance of Jovian and Neptune mass planets is one power of ten more than that predicted by the formation models around M-dwarf stars (Suzuki et al. in prep.).
However, as the hydrogen and helium in their protoplanetary disks would be lost within several million years, massive Jovian planets are unlikely to be formed around VLMS and BDs because there would, as a result, be insufficient materials to form gas giants (\citealp{2005ApJ...626.1045I}).
\vspace{0.2in}

Without an observed microlensing parallax, we cannot determine the absolute masses of the lens system from a microlensing light curve alone. 
In such cases, a Bayesian analysis is usually applied by using Galactic model priors (comprising a mass function and number density, and velocity distributions) to obtain the posterior probability distributions of the physical parameters of the lens system. 
The results of the Bayesian analysis generally strongly depend on the priors. 
In addition, we should pay special attention to low mass hosts because the mass function for the low mass region ranging from BDs down to planetary mass objects is still very uncertain.
Gravitational microlensing plays an important role for probing the occurrence of planets in distant orbits and planetary mass objects unbound to any host stars. 
By investigating the time scale distribution of short microlensing events in the MOA dataset in 2006-2007, \citet{2011Natur.473..349S} suggested a possible population excess of unbound or distant Jupiter-mass objects.
Unbound planets are thought to derive from various physical processes.
There are, for instance, star-planet scattering (\citealp{1999AJ....117..621H}; \citealp{2005A&A...434..355M}; \citealp{2011MNRAS.418.2656D}; \citealp{2011MNRAS.411..859M};  \citealp{2012MNRAS.421L.117V}; \citealp{2013Natur.493..381K}), planet-planet scattering (\citealp{1998AJ....116.1998L}; \citealp{2008ApJ...686..621F}; \citealp{2011ApJ...732...74G}), and  stellar mass loss (\citealp{2011MNRAS.417.2104V}; \citealp{2012MNRAS.422.1648V}; \citealp{2013MNRAS.430.3383V}). 
\citet{2017Natur.548..183M} recently updated the study with a larger sample and reported no significant excess of short-timescale (1-2 days) microlensing events and placed $95\%$ upper limits on the frequency of unbound or distant Jupiter mass objects of 0.25 planet per main sequence star using the OGLE 2010-2015 observations. 
They also reported a possible abundance of unbound or distant super-Earth mass objects (\citealp{2018ApJ...853L..27D}). 
These studies have played an important role for probing the mass function for low mass objects, while it is still uncertain. 
\vspace{0.2in}

In order to find as many microlensing events as possible, a large number of stars must be monitored because the probability of any given star being microlensed is very low.
The second phase of the Microlensing Observations in Astrophysics (MOA; \citealp{2001MNRAS.327..868B};  \citealp{2003ApJ...591..204S}) collaboration, MOA-II, has continued microlensing survey observations using the 1.8 m MOA-II telescope with a 2.2 deg$^2$ wide field of view (FOV) CCD camera (MOA-cam3; \citealp{2008ExA....22...51S}) at the Mount John Observatory (MJO) in New Zealand.
The MOA observational fields are mainly toward Baade's window in the Galactic bulge and they observe the fields with cadences of $\sim$10-50 times per night.
Six MOA-II fields ($\sim$13 deg$^2$) are observed with a 15 minute cadence, seeking particularly short-time microlensing events.
About 600 microlensing event alerts are issued by MOA-II in real time each year.\footnote{https://www.massey.ac.nz/~iabond/moa/alerts/}
Other microlensing survey groups are the Optical Gravitational Lensing Experiment (OGLE; \citealp{2015AcA....65....1U}) collaboration and the Korea Microlensing Telescope Network (KMTNet; \citealp{2016JKAS...49...37K}).  
About 2000 microlensing event alerts are issued by OGLE in real time each year\footnote{http://ogle.astrouw.edu.pl/ogle4/ews/ews.html}, and about 2200 alerts are issued by KMTNet each year\footnote{http://kmtnet.kasi.re.kr/ulens/}.
\vspace{0.2in}

In this paper, we present the analysis of the short-timescale microlensing event, MOA-2015-BLG-337.
We find two competing models that explain the observed data.
One comprises a planetary mass-ratio lens system and the other, a binary mass ratio lens system.
Because of the short time scale, $\sim 6$ days, of this event, we conducted our Bayesian analysis with an assumed mass function extending down to brown dwarf and also down to planetary mass regimes.
We describe the observations of microlensing and the data sets for this event in Section \ref{sec:obs}.
Our light curve modeling is explained in Section \ref{sec:lightcurvemodel}.
In Section \ref{sec:AER}, we present the calibration of the source star and the estimation of the angular Einstein radius.
In Section \ref{sec:Bayesian}, we perform a Bayesian analysis in order to determine the probability distribution of the physical parameters of the lens system.
In Section \ref{sec:Discussion}, we discuss the results of this work.
Finally, we present the summary and our conclusion in Section \ref{sec:SAC}.
}
 
\section{Observation} \label{sec:obs}
\begin{figure}[htb!]
\centering
\includegraphics[scale=0.55, angle=-90]{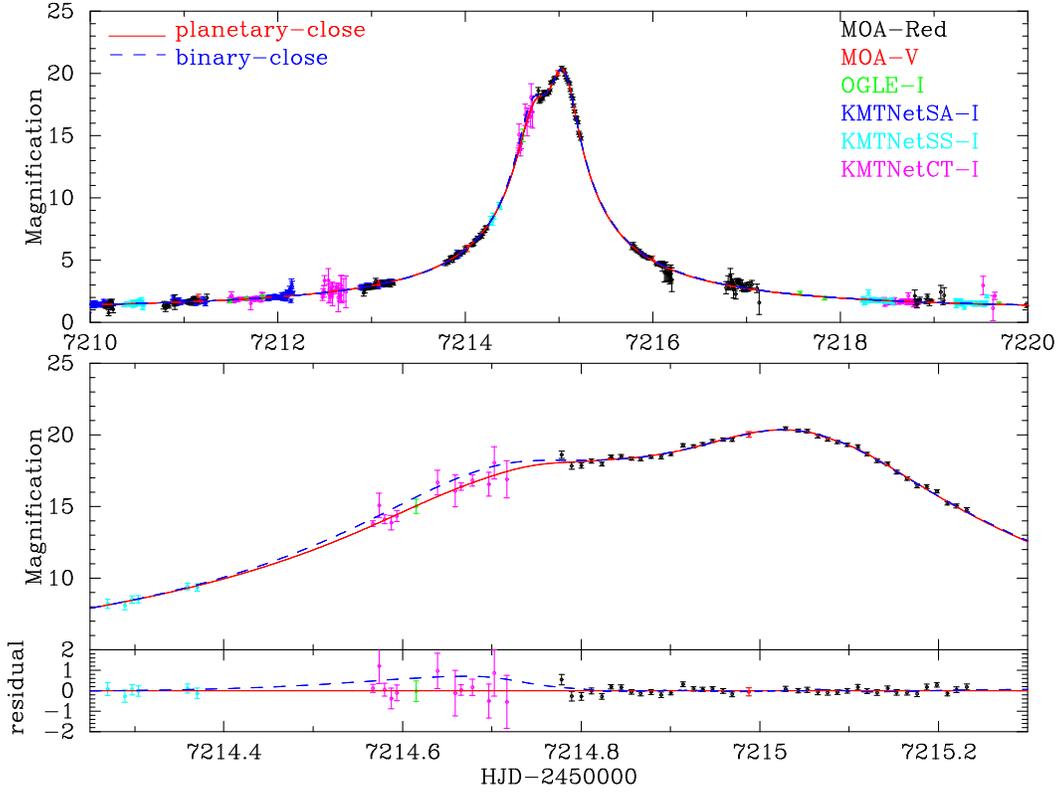}
\caption{
The light curve of MOA-2015-BLG-337. 
The top, middle and bottom panels show the entire period of the event, close-up of the anomaly and the residual from the best fitting planetary-close model, respectively. 
The red solid line and blue dashed line indicate the best planetary-close model and the best binary-close model.  
The binary model deviates from the planetary model particularly between $7214.5 < {\rm HJD}-2450000 < 7214.8$.
The large photometric uncertainties in this period make it difficult to distinguish which is the best-fit model.
The B\&C data are not shown for clarity because of their large error bars while these data are used in the analysis.
}
\label{fig:light-curve}
\end{figure}

\normalsize{
Event MOA-2015-BLG-337 was first discovered by MOA on 2015 July 10 UT (${\rm HJD}'\equiv{\rm HJD}-2450000=7214$) and positioned at $({\rm RA}, {\rm Dec})_{\rm J2000}=(18^{\rm h}07^{\rm m}47^{\rm s}.69, -28^{\circ}10^{'}13^{''}.00)$, which corresponds to Galactic coordinates $(l, b) = (3.11^{\circ}, -3.83^{\circ}$).
The event is located in MOA field ``gb14'' which is observed with a 15 minute cadence.
OGLE independently found and alerted the event as OGLE-2015-BLG-1598 on 2015 July 11 UT (${\rm HJD}'=7215$) with their 1.3 m Warsaw telescope with a 1.4 deg$^{2}$ FOV at LCO. 
KMTNet also observed this event as KMT-2015-BLG-0511 (\citealp{2018AJ....155...76K}) in their regular observation survey with their three 1.6 m KMTNet telescopes each having a 4.0 deg$^{2}$ FOV at CTIO, SAAO and SSO.
\vspace{0.2in}

The MOA observers noticed that the light curve of the MOA-2015-BLG-337 deviated from a single-lens model around HJD$^{'} \sim 7214.9$ and the MOA collaboration issued an anomaly alert encouraging follow-up observations. 
Some modelers in microlensing survey and follow-up groups immediately modeled this event and found two indistinguishable solutions with planetary and binary lens mass ratios.
KMTNet-CTIO and OGLE  light curves also cover the part of the anomaly. 
This event was also observed by the 0.61 m Boller $\&$ Chivens telescope at MJUO. 
The number of data points of each telescope and passband are shown in Table \ref{table:data}. 
The MOA and B$\&$C data were reduced with MOA's photometry pipeline (\citealp{2001MNRAS.327..868B}) which uses the Difference Image Analysis (DIA) technique (\citealp{1996AJ....112.2872T}; \citealp{1998ApJ...503..325A}; \citealp{2000A&AS..144..363A}).
The OGLE data were reduced with the OGLE DIA (\citealp{2000AcA....50..421W}) photometry pipeline (\citealp{2015AcA....65....1U}).
The KMTNet data also were reduced with their PySIS photometry pipeline (\citealp{2009MNRAS.397.2099A}; \citealp{2016JKAS...49...37K}).
\vspace{0.2in}

The photometric errors estimated from pipelines are generally underestimated owing to the high stellar densities in the Galactic bulge.
We, therefore, renormalized the error bars of each dataset following the method described in \citet{2012ApJ...755..102Y}, i.e., $\sigma^{'}_{i}=k\sqrt{\sigma^{2}_{i}+e^{2}_{\rm min}}$, where $\sigma^{'}_{i}$ and $\sigma_{i}$ are the renormalized and original  error bar of the $i$-th data points in magnitude, $k$ and $e_{\rm min}$ are parameters for renormalizing.
The value of $e_{\rm min}$ represents systematic errors that dominates at high magnification and can be affected by flat-fielding errors. 
First, we fit the light curve to find a tentative best-fit model.
Then, we fitted the $k$ and $e_{\rm min}$ values so that the cumulative $\chi^{2}$ distribution from the tentative best model sorted by model-magnification is $\chi^{2}/d.o.f\sim1$ and close to linear with a slope of 1, respectively. 
However, we could not constrain $e_{\rm min}$ in most of datasets because our datasets are not highly magnified, thus not sensitive to $e_{\rm min}$.
So, we set $e_{\rm min}=0.003$, which is empirically chosen, and selected $k$ value in order to be $\chi^{2}/d.o.f\sim1$ as shown Table \ref{table:data}.
The renormalizing parameters applied for the OGLE-{\it I} data are consistent with those taken from \citet{2016AcA....66....1S}.
After the error renormalization, we fitted the light curve again to find the final best model.
In general, error renormalization processes do not significantly affect the final results.
We confirmed the final best model is consistent with the previous tentative best-fit model before error renormalization.
\vspace{0.2in}

Figure \ref{fig:light-curve} represents the light curve of this event. 
It shows a clear asymmetric feature, which cannot be explained by a finite-source point-lens (FSPL) model by $\Delta\chi^{2} > 1600$.
We describe the procedure of binary-lens modeling in the following section.
}

\section{Light Curve Models} \label{sec:lightcurvemodel}
\subsection{Model Description}
\normalsize{
The magnification of a standard binary lens model has seven parameters; the time of the source closest to the center of mass, $t_{0}$; the Einstein radius crossing time, $t_{\rm E}$; the impact parameter normalized by the Einstein radius, $R_{\rm E}$, $u_{0}$; the mass ratio of a lens companion relative to a host, $q$; the projected separation normalized by $R_{\rm E}$ between binary components, $s$; the angle between the binary lens axis and the source trajectory direction, $\alpha$; and finally the source angular radius relative to the angular Einstein radius, $\theta_{\rm E}$, $\rho$.
If $\rho$ is measured from the light curve modeling, we can estimate the angular Einstein radius, $\theta_{\rm E} = \theta_{*}/\rho$, because the source angular radius, $\theta_{*}$, can be estimated from its apparent magnitude and color corrected for extinction and reddening (\citealp{2014AJ....147...47B}).
Given parameters ${\bf x} = (t_{\rm E}, t_{0}, u_{0}, q, s, \alpha, \rho)$, the magnification $A(t, {\bf x})$ can be calculated at any given time, $t$.
The model light curve can be written as :
\begin{equation}
F(t) = A(t, {\bf x})f_{\rm s} + f_{\rm b},
\end{equation}
where $f_{\rm s}$ is the un-magnified source flux, and $f_{\rm b}$ is the blend flux. 
Each telescope and pass-band has a corresponding $f_{\rm s}$, $f_{\rm b}$ pair.
We applied a linear limb-darkening model for the source, $I(\theta)=I(0)[1-u_{\lambda}(1-$cos$\theta)]$.
Here, $\theta$ and $u_{\lambda}$ represent the angle between the line of sight and the normal to the stellar surface, and the linear limb-darkening coefficient, respectively.
We estimated the effective temperature of the source, $T_{\rm eff}\sim $ 6000K, and assumed a metallicity, log[$M/H$] = 0.0 and surface gravity ${\rm log}(g/{\rm cm\;s^{-1}})=4.50$. 
According to the ATLAS model of \citet{2011A&A...529A..75C}, the limb-darkening coefficients are $u_{R}=0.7021, u_{I}=0.6055$ and $u_{V}=0.7801$.
The $u_{R+I}$ value, corresponding to limb darkening in the MOA-$Red$ passband, is estimated as the mean of $u_{R}$ and $u_{I}$.
}

\subsection{Modeling}
\begin{figure}[htb!]
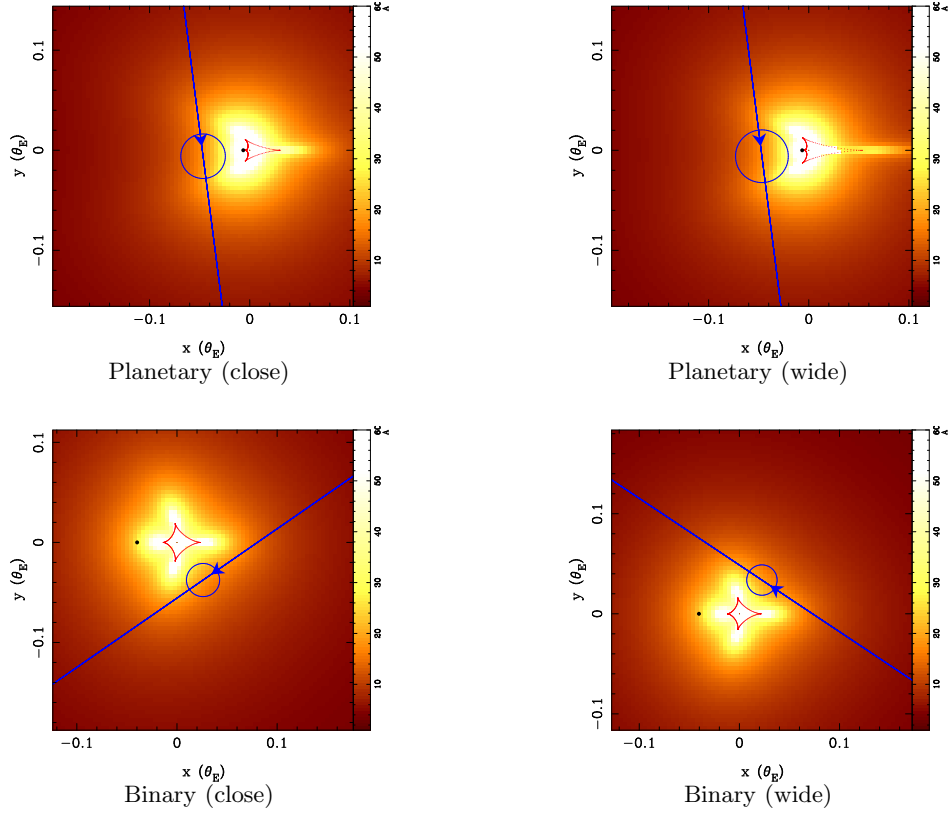

  \begin{center}
    \begin{tabular}{c}
      \begin{minipage}{0.4\hsize}
        \begin{center}
          \includegraphics[keepaspectratio,scale=0.24, angle=270]{./PLC-cau.ps}
          \hspace{1.8cm} Planetary (close) 
        \end{center}
      \end{minipage}
      \begin{minipage}{0.4\hsize}
        \begin{center}
          \includegraphics[keepaspectratio,scale=0.24, angle=270]{./PLW-cau.ps}
          \hspace{1.8cm} Planetary (wide) 
        \end{center}
      \end{minipage}
      \\
      \\
      \begin{minipage}{0.4\hsize}
        \begin{center}
          \includegraphics[keepaspectratio,scale=0.24, angle=270]{./BIC-cau.ps}
          \hspace{1.8cm} Binary (close)
        \end{center}
      \end{minipage}
      \begin{minipage}{0.4\hsize}
        \begin{center}
          \includegraphics[keepaspectratio,scale=0.24, angle=270]{./BIW-cau.ps}
          \hspace{1.8cm} Binary (wide)
        \end{center}
      \end{minipage}
    \end{tabular}
    \caption{
    Caustic geometries for each model are shown as the red curves, respectively.
    The magnification patterns around them are also represented as color maps.
    The brighter tone denotes higher magnifications.
    The blue lines show the source star trajectory with respect to the lens system, with the arrows indicating the direction of motion.
    The blue circles on the lines indicate the size of the source. 
    The black dots represent the positions of the lens hosts.
    }
    \label{fig:caustic}
  \end{center}
\end{figure}

\normalsize{
We used a Markov Chain Monte Carlo (MCMC) approach (\citealp{2003ApJS..148..195V}) in order to search for the best fit model parameters combined with our implementation of the image ray-shooting method (\citealp{1996ApJ...472..660B}; \citealp{2010ApJ...716.1408B}).
At first, we conducted a broad grid search by using the standard static binary lens model with 9,680 fixed grid points covering wide ranges of the three parameters, $q$, $s$ and $\alpha$, with all other parameters allowed to vary. 
Then, the best 100 models with the smallest $\chi^{2}$ were refined with all free parameters.
With these procedures, we can find the best fit model and avoid missing any local minima in the wide range of the parameters space.
After that, we found a pair of optimal solutions, one with a planetary mass ratio of $q\sim1.1\times10^{-2}$ and the other with a stellar binary mass ratio of $q\sim1.7\times10^{-1}$. 
These parameters are labeled as $``$Planetary$"$ and $``$Binary$"$ in Table \ref{model-parameter}, respectively.
We could not find any model with a significant microlensing parallax effect as expected because of the short event time scale.
In addition, we examined the possibility of a binary-source point-lens (BSPL) model for this event, and confirmed it could be ruled out by $\Delta\chi^{2} > 50$.
\vspace{0.2in}

In the planetary model, there are two statistically indistinguishable solutions, one with a close ($s<1$) binary separation and the other with a wide ($s>1$) separation.
The close model is favored by only $\Delta\chi^{2}\simeq0.2$.
This severe degeneracy between $s$ and $s^{-1}$, which is known as close/wide degeneracy, is common for central caustic crossing events because the shapes of the central-caustic with $s$ and $s^{-1}$ are similar for smaller $q$ (\citealp{griest98}; \citealp{1999A&A...349..108D}).
In these planetary models, the finite source effect with $\rho \sim 2\times10^{-2}$  is detected with $\Delta\chi^{2} > 35$ compared to the point source models, and the signal mostly came from MOA-$Red$ data.
This $\rho$ value is relatively large compared to the typical values for other microlensing planetary events and it indicates a small angular Einstein radius, $\theta_{\rm E}$, and prefers a smaller lens mass in the Bayesian analysis of Section 5. 
\vspace{0.2in}

In the binary model, the model deviates from the planetary model particularly between $7214.5 < {\rm HJD}-2450000 < 7214.8$ (see Figure \ref{fig:light-curve}).
The difference between these models is only $\Delta\chi^{2}\sim6$ due to the large photometric uncertainties in this period. 
There is also a close/wide degeneracy in the binary model: the close model is favored over the wide one by $\Delta\chi^{2}\sim2$.
The finite source effect was not significantly detected for these binary models, showing only an improvement of $\chi^{2} < 4.0$ compared to the point source model.
Thus, we do not adopt the best-fit $\rho$ value but put only an upper limit on $\rho$ in the following analysis.
 \vspace{0.2in}
 
Our inability to distinguish between the planetary and binary models is owing to the lack of data during the anomaly where there are only subtle differences in the magnification pattern between these models.
Figure \ref{fig:caustic} represents the caustic geometries and the magnification patterns around them for each model.
The magnification pattern this event can be explained by the two different central caustic shapes, which makes it difficult to distinguish the two planetary and stellar binary solutions.
Similar to this event, there are a number of microlensing events which can be explained with both stellar binary and planetary mass ratio lens systems (\citealp{2004ApJ...611..528G}; \citealp{2012ApJ...756...48C}; \citealp{2014ApJ...787...71P}).
}

\section{Angular Einstein  Radius}\label{sec:AER}
\begin{figure}[htb!]
\centering
\includegraphics[scale=0.5, angle=-90]{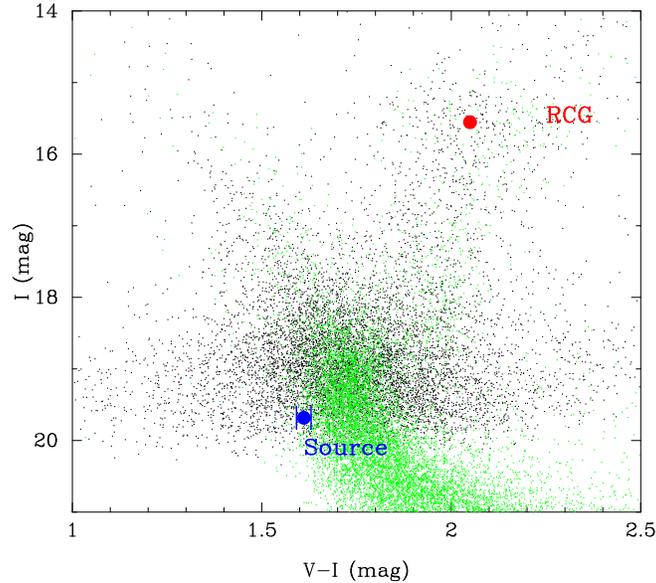}
\caption{
Color Magnitude Diagram (CMD) of OGLE-III stars within 2' of MOA-2015-BLG-337 (Black dots). 
The green dots indicate the Hubble Space Telescope (HST) CMD in Baade's window (\citealp{1998AJ....115.1946H}) whose color and magnitude are matched by using the RCG position. 
The red point indicates the centroid of red clump giant in this field, and the blue point indicates the source star in this event. 
}
\label{fig:cmd}
\end{figure}

\normalsize{
By using the measurement of $\rho$ $(=\theta_{*}/\theta_{\rm E})$ from the models, the source angular radius $\theta_{*}$ allows us to estimate $\theta_{\rm E}$. 
We can estimate $\theta_{*}$ from the source intrinsic color and magnitude (\citealp{2014AJ....147...47B}).
\vspace{0.2in}

We converted the instrumental source magnitude in MOA-{\it Red} and MOA-{\it V} bands into the standard Kron-Cousin {\it I}-band and Johnson {\it V}-band scales using the model light curve.
With the procedure described in \citet{bond17}, we cross-referenced stars within 2' of the source between the MOA star catalog reduced by DoPHOT (\citealp{1993PASP..105.1342S}) and the OGLE-III star catalog (\citealp{2011AcA....61...83S}).
Then, we found the following relations,
\begin{eqnarray*}
	I_{\rm OGLE-III}-R_{\rm MOA} & = & (28.119\pm0.005)-(0.206\pm0.002)(V-R)_{\rm MOA}\\
	V_{\rm OGLE-III}-V_{\rm MOA} & = & (27.901\pm0.005)-(0.148\pm0.002)(V-R)_{\rm MOA}.
\end{eqnarray*}
We tried to correct interstellar extinction and reddening by following the standard procedure in \citet{2004ApJ...603..139Y}, which treats the Red Clump Giants (RCG)  in the color-magnitude diagrams (CMD) as standard candles.
Figure \ref{fig:cmd} shows the CMD of the OGLE-III star catalog within 2' of the source, plotted over the CMD of the Baade's window from \citet{1998AJ....115.1946H} whose extinction and reddening are matched by using RCG position.
The magnitude and color of source and the center of the RCG, $(V-I, I)_{\rm s}=(1.611, 19.682)\pm(0.019, 0.012)$ and $(V-I, I)_{\rm RCG}=(2.050, 15.553)\pm(0.013, 0.048)$ are shown as filled blue and red circles, respectively.
We obtained the intrinsic RCG centroid in this field $(V-I, I)_{\rm RCG,0}=(1.060, 14.348)\pm(0.070, 0.040)$ (\citealp{2013A&A...549A.147B}; \citealp{2013ApJ...769...88N}), and found the extinction and reddening of RCG centroid to be $A_{I,{\rm RCG}}=1.205\pm0.062$ and $E(V-I)_{\rm RCG}=0.990\pm0.071$, respectively.
The intrinsic  source magnitude and color values are $(V-I, I)_{\rm s,0}=(0.621, 18.478)\pm(0.074, 0.064)$ on the assumption that the source star suffers from the same extinction and reddening as that of the bulge RCGs in the field.
From the CMD, we can see that the source star is slightly bluer than other typical bulge dwarf stars in this field. 
This is possibly because the source does not have the same extinction and reddening as the median values of the RCG stars.
However, even if we assume 10\% less extinction and reddening from those for the median RCG,  the changes in the $\theta_{\rm E}$ and $\mu_{\rm rel}$ are less than 3\%, which is much less than the typical uncertainty of these values.
We summarize the source color, magnitude and angular radius in Table \ref{tab:source}.
We also independently obtained the intrinsic source color using a linear regression from KMTNet-CTIO $I$ and $V$, $(V-I)_{\rm s,0,KMTNet}=0.688\pm0.066$, which is consistent with the color from MOA-$Red$ and $V$ above but with a smaller uncertainty. 
We applied this $(V-I)_{\rm s,0,KMTNet}$ in the following analysis.
\vspace{0.2in}

For estimation of the source angular radius $\theta_{*}$,  we used the following empirical relation, 
\begin{equation}
{\rm log}\theta_{\rm LD} = 0.5014 + 0.4197(V-I) - 0.2I ,
\end{equation}
where $\theta_{\rm LD}\equiv2\theta_{*}$ (\citealp{2015ApJ...809...74F}).
This empirical relation is derived by using a subsample of FGK stars between $3900{\rm K}<T_{\rm eff}<7000{\rm K}$ from \citet{2014AJ....147...47B} and the accuracy of this relation is better than 2\%.
According to this relation, we found that $\theta_{*}\equiv\theta_{\rm LD}/2=0.621\pm0.045\;\mu{\rm as}$, and angular Einstein radius $\theta_{\rm E}$ and the lens-source relative proper motion $\mu_{\rm rel}$ are given as,
\begin{eqnarray*}
\theta_{\rm E}=\frac{\theta_{*}}{\rho} = \begin{cases} 
					 0.028\pm0.004\;{\rm mas} \;\rm{(for\;the\;planetary\;close\;model)} \\
                                   0.024\pm0.003\;{\rm mas\;(for\;the\;planetary\;wide\;model)} \\
                                    >  0.034\;{\rm mas}\;{\rm (for\;the\;binary\;close\;model)} \\
                                    >  0.035 \;{\rm mas}\;{\rm (for\;the\;binary\;wide\;model)} \\
                                      \end{cases}\\
\mu_{\rm rel}=\frac{\theta_{\rm E}}{t_{\rm E}} =
					 \begin{cases} 
					 1.90\pm0.29\;{\rm mas\;yr^{-1}\;(for\;the\;planetary\;close\;model)} \\
                                    1.59\pm0.18\;{\rm mas\;yr^{-1}\;(for\;the\;planetary\;wide\;model)}\\
                                    >  2.26\;{\rm mas\;yr^{-1}\;(for\;the\;binary\;close\;model)} \\ 
                                    >  2.11\;{\rm mas\;yr^{-1}\;(for\;the\;binary\;wide\;model)}  ,\\
                                    \end{cases}
\end{eqnarray*}
where only the lower limits are given for the binary models.
}

\section{Bayesian analysis}\label{sec:Bayesian}
\begin{figure}[htb!]
\centering
\includegraphics[scale=0.8,angle=-90]{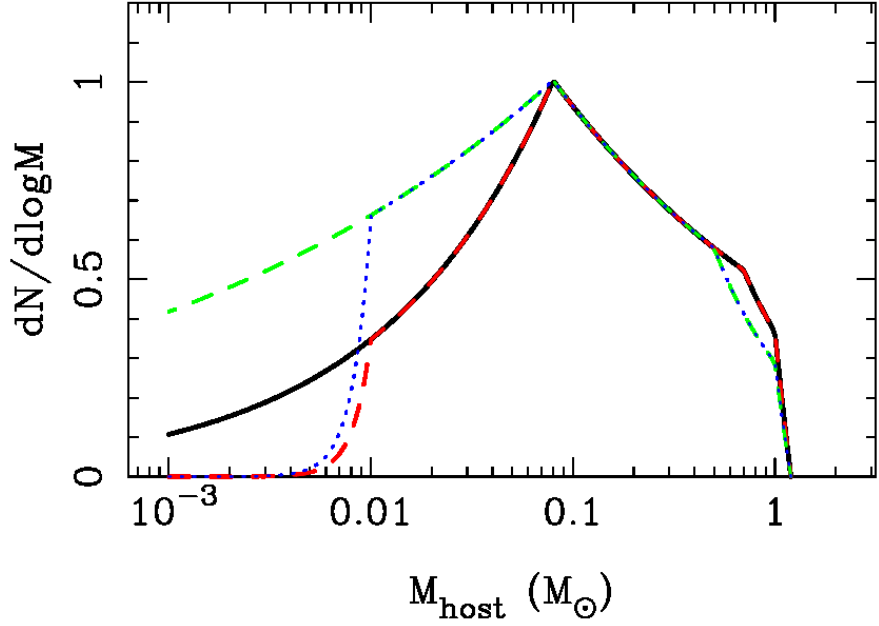}
\caption{
The mass functions assumed in the Bayesian analysis in Section 5, and equation (\ref{eq:power-law}).
The black solid, the red dashed, the green dashed and the blue dotted lines represent those with $(\alpha_{\rm bd}, \alpha_{\rm pl})=(0.49,0.49), (0.49, -4.0), (0.8,0.8)$ and $(0.8, -4.0)$, respectively.
In these mass functions, the abundance ratios of main sequence stars, brown dwarfs and planetary objects are 1:0.72:0.27, 1:0.72:0.04, 1:0.99:0.70 and 1:0.99:0.07.
}
\label{fig:imf}
\end{figure}

\begin{figure}[htb!]
\centering
\includegraphics[scale=0.45, angle=-90]{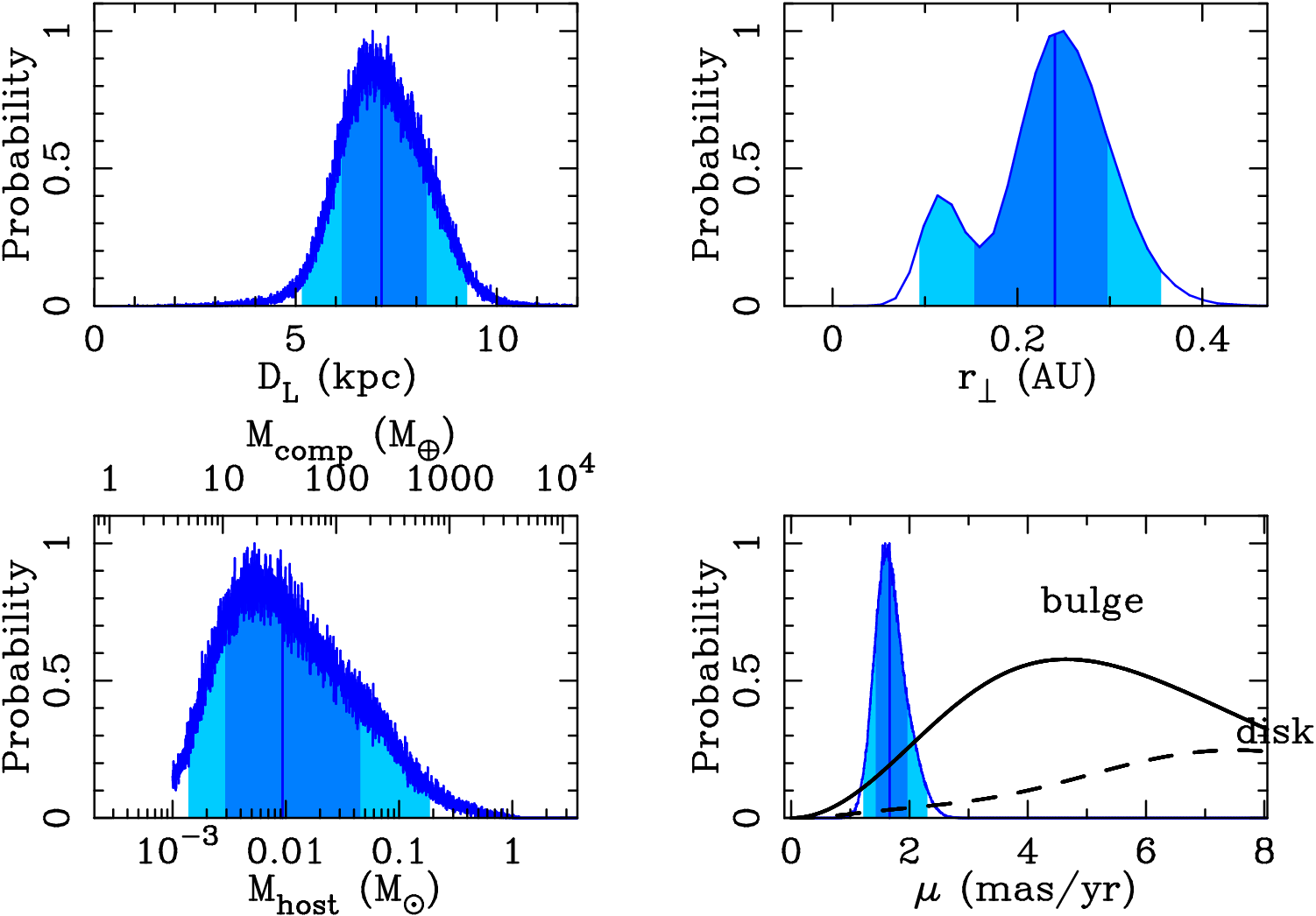}
\caption{
Probability distribution of lens properties for the {\bf planetary} model derived from the Bayesian analysis with applying equation (\ref{eq:power-law}) with $\alpha_{\rm bd}=0.49$, $M_{\rm break}=0.7$ and  ${\alpha_{\rm pl}=0.49}$.
The dark and light blue regions indicate the 68.3\% and 95.4\% confidence interval, and the vertical blue lines indicates the median value.
The right bottom panel shows the probability distribution for the lens-source relative proper motion $\mu$, and the black line and the black dashed line indicate the prior $\mu$ probability distribution for Galactic bulge and disk, derived from the Galactic model (\citealp{1995ApJ...447...53H})
}
\label{fig:bayesian-plc-Sumi-extend}
\end{figure}

\begin{figure}[htb!]
\centering
\includegraphics[scale=0.47, angle=-90]{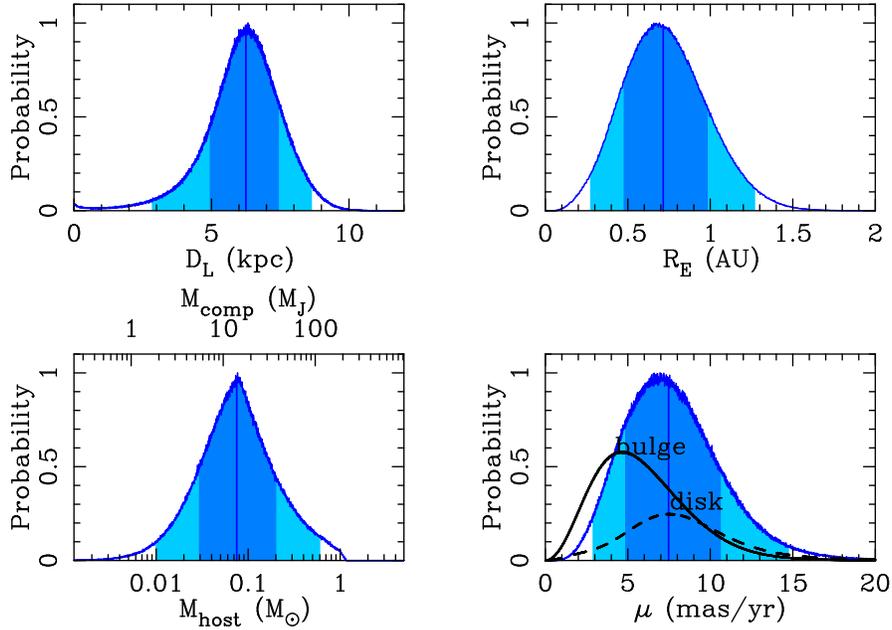}
\caption{
Probability distribution of lens properties for the {\bf binary} model derived from the Bayesian analysis with applying equation (\ref{eq:power-law}) with $\alpha_{\rm bd}=0.49$, $M_{\rm break}=0.7$ and $\alpha_{\rm pl}=0.49$.
The dark and light blue regions indicate the 68.3\% and 95.4\% confidence interval, and the vertical blue lines indicates the median value.
The right bottom panel shows the probability distribution for the lens-source proper motion $\mu$, and the black line and the black dashed line indicate the prior $\mu$ probability distribution for Galactic bulge and disk, derived from the Galactic model (\citealp{1995ApJ...447...53H}) .
}
\label{fig:bayesian-bin-Sumi-extend}
\end{figure}

\normalsize{
Since no significant parallax signal is detected in this event, we cannot directly measure the lens properties, the host mass, $M_{\rm host}$, the distance, $D_{\rm L}$ or lens-source relative transverse velocity, $v_{\perp}$.
We, therefore, performed a Bayesian approach to explore the probability distribution of the lens characteristics (\citealp{2006ApJ...644L..37G}; \citealp{2006Natur.439..437B}; \citealp{2008ApJ...684..663B}).
We assumed the Galactic model (\citealp{1995ApJ...447...53H}) as the priors for Galactic mass density and velocity.
The observed $t_{\rm E}$ and $\theta_{\rm E}$ can constrain the lens physical parameters in the Bayesian analysis.
Both the observed $t_{\rm E}$ and $\theta_{\rm E}$ value in this event are smaller than average.
This suggests that the host mass is likely to be very low because both $t_{\rm E}$ and $\theta_{\rm E}$ are proportional to $\sqrt{M}$.
Therefore, we need a mass function extending to a very low mass regime in the Bayesian analysis.  
We applied the broken power-law mass function used  in \citet{2011Natur.473..349S} and \citet{2017Natur.548..183M} as follows,
\begin{eqnarray}
\label{eq:power-law}
dN/dM= \begin{cases}
a_{0}M^{-\alpha_{\rm pl}}       & (0.001 \le M/M_{\odot} \le 0.01)\\
a_{1}M^{-\alpha_{\rm bd}}      & (0.01 \le M/M_{\odot} \le 0.08) \\
a_{2} M^{-\alpha_{\rm ms2}}   & (0.08 \le M/M_{\odot} \le M_{\rm break}) \\
a_{3}M^{-\alpha_{\rm ms1}}    & (M_{\rm break} \le M/M_{\odot}\le 1.0) ,
\end{cases}
\end{eqnarray}
where $\alpha_{\rm ms1}=2.0$ and $\alpha_{\rm ms2}=1.3$ are the power-law indexes for main sequence stars. 
We adopted the power-law index for the brown dwarfs of $\alpha_{\rm bd}=0.49$ with $M_{\rm break}=0.7$ (\citealp{2011Natur.473..349S}) and $\alpha_{\rm bd}=0.8$ with $M_{\rm break}=0.5$ (\citealp{2017Natur.548..183M}).
\citet{2017Natur.548..183M} assumed the low mass end of the mass function of $M/M_{\odot}=0.01$ and found that the upper limit of the abundance of Jupiter mass objects is 0.25 per star with 95\% confidence.
We applied the case for extending the brown dwarf's slope to the planetary mass regime with $\alpha_{\rm bd}=\alpha_{\rm pl}$, and the case with sharp decline below $0.01M_{\odot}$ with $\alpha_{\rm pl}=-4.0$.
The mass functions with $\alpha_{\rm pl}=-4.0$ are similar to those in \citet{2017Natur.548..183M} but with slightly gentle cutoff.
Figure \ref{fig:imf} shows these mass functions.
In these mass functions, the relative fractions of number densities between main sequence stars, brown dwarfs and planetary mass objects for ($\alpha_{\rm bd}$, $\alpha_{\rm pl}$) = $(0.49, 0.49)$, $(0.49, -4.0)$, $(0.8, 0.8)$ and $(0.8, -4.0)$ are $1:0.72:0.27$, $1:0.72:0.04$, $1:0.99:0.70$ and $1:0.99:0.07$, respectively.
Here the model with $(\alpha_{\rm bd}, \alpha_{\rm pl})=(0.49,0.49)$ has 0.27 planetary mass objects per main sequence stars, which is just slightly higher than the 95\% upper limit of Jupiter-mass planetary mass objects by \citet{2017Natur.548..183M}.
Thus, the abundance of planetary-mass objects in this model can be considered as an upper limit. 
The model with $(\alpha_{\rm bd} , \alpha_{\rm pl})=(0.8,0.8)$ has even more planetary-mass objects, thus it can be considered as an extreme case.
On the other hand, the models with $(\alpha_{\rm bd} , \alpha_{\rm pl})=(0.49, -4.0)$ and $(0.8, -4.0)$ can be considered as reference cases if we assume much lower  abundance of planetary mass objects than the upper limit of 0.25 per star. 
In this analysis, we omitted the mass functions of stellar remnants because the host mass is too low to be affected.
Here, the number of main sequence stars decreases linearly to zero from $1M_{\odot}$ to $1.2M_{\odot}$ to avoid the unphysically sharp cutoff at $1M_{\odot}$.
We assumed main sequence stars with masses above 1.2 $M_{\odot}$ have developed into stellar remnants due to their short lifetime (\citealp{2011Natur.473..349S}).
We also assumed that the probability of the lens star hosting a companion with the measured physical parameters is independent of the lens host mass because we don't have any information about that for very low mass hosts.
\vspace{0.2in}

The lens parameters and the probability distributions of the lens properties derived from our Bayesian analysis are given in Tables \ref{tab:Sumi-IMF} and \ref{tab:Mroz-IMF} and Figures \ref{fig:bayesian-plc-Sumi-extend} and \ref{fig:bayesian-bin-Sumi-extend}.
We combined the probability distribution of the close and wide models by weighting the probabilities of the wide models parameters by $e^{-\Delta\chi^{2}/2}$, in which $\Delta\chi^{2}=\chi^{2}_{\rm wide}-\chi^{2}_{\rm close}$.
However, the probability distribution of the projected separation between the host and the companion $r_{\perp}$ for the binary model is not combined because the difference of the $s$ values between the close and wide models are so large, $s_{\rm wide}/s_{\rm close} \simeq 18$, compared to that value of the planetary models of $\simeq 2.5$.  
\vspace{0.2in}

As one can see in these tables, the results do not depend strongly on the value of $\alpha_{\rm bd}$.
However, the host mass for the planetary model strongly depends on the assumed $\alpha_{\rm pl}$.
For the planetary models with $\alpha_{\rm pl}$ of $-4.0$, the lens system is likely a brown dwarf orbited by a Saturn mass giant planet.  
The lens masses for the planetary models  with $\alpha_{\rm pl}=0.49$ and 0.8 are lower than that with $\alpha_{\rm pl}=-4.0$ as expected, but consistent to within 1$\sigma$.
This is because of the strong restrictions from the observed values of $t_{\rm E}$ and $\theta_{\rm E}$.
In the planetary models with $\alpha_{\rm pl}=0.49$, the host, which is on the boundary between a super giant planetary mass object and a very low-mass brown dwarf, is orbited by a super Neptune-mass planet or a sub-Saturn-mass planet.
The assumption of $\alpha_{\rm pl}=-4.0$, i.e., the sharp decline  in the lens mass function below $0.01 M_{\odot}$ seems unphysical because there are many claims of brown dwarfs with masses below $0.01 M_{\odot}$. 
Thus, we do not consider the models with $\alpha_{\rm pl}=-4.0$ as a final result but as a reference for comparison.
For the binary model, on the other hand, the dependency of the Bayesian results on $\alpha_{\rm bd}$ and $\alpha_{\rm pl}$ is not strong. 
This is because we only have a lower limit on $\theta_{\rm E}$, thus a relatively more massive host is allowed in the Bayesian analysis.  
This model indicates that the lens system is a brown dwarf binary with a small mass ratio. 
In addition, we conducted the independent Bayesian analysis with a different Galactic model by using the same procedure as \citet{2014ApJ...785..155B}. 
We confirmed that the results of the Bayesian analysis were consistent with those in Table \ref{tab:Sumi-IMF} and \ref{tab:Mroz-IMF}.
We found that the derived lens-source proper motion, $\mu$, for the binary solutions is preferred than that of the planetary solutions by the Galactic model prior. 
However, this preference is not so large compared to the preference of the planetary model by the $\Delta\chi^{2}\sim6$ from the light curve fitting.
So, we concluded that this event has an ambiguity between two solutions while the planetary solution is slightly preferred.
}

\section{Discussion}\label{sec:Discussion}
\begin{figure}[htb!]
\centering
\includegraphics[scale=0.47, angle=-90]{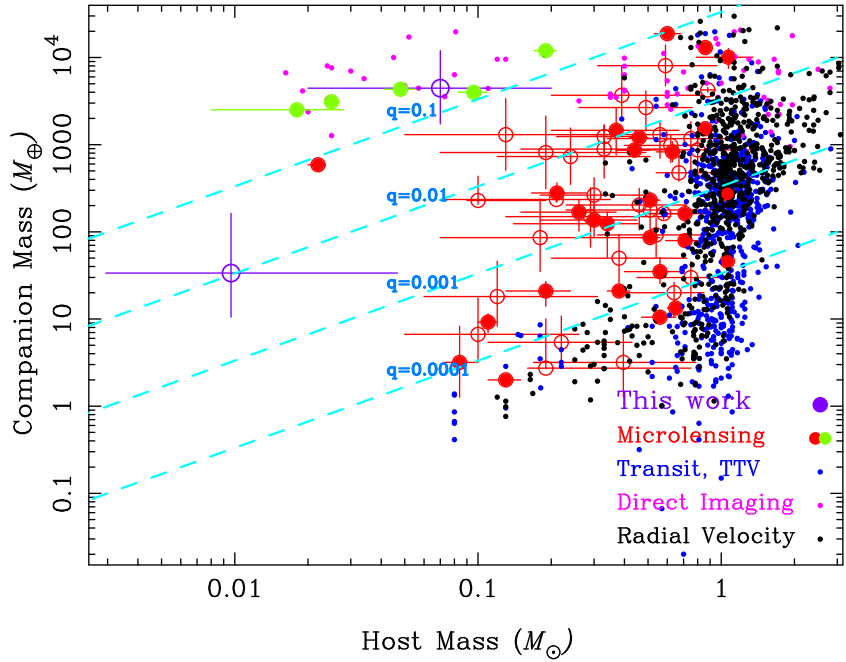}
\caption{
Distribution of bound exoplanets and very low mass companions in which the vertical axis shows the companion masses and the horizontal axis shows the host masses.
The two purple points indicates the results of the Bayesian analysis for MOA-2015-BLG-337.
The red, green, blue, magenta and black points indicate the planetary systems found by Microlensing (with a mass ratio of planet/host of $q<0.1$), Microlensing ($q>0.1$), Transit $\&$ TTV, Direct Imaging, and Radial Velocity, respectively. 
For the microlensing planets, filled circles indicate that their masses are measured and open circles indicate that their masses are estimated by a Bayesian analysis.
The values of microlensing planets are from each discovery paper, while those of the others are from http://exoplanet.eu.
}
\label{fig:exoplanet}
\end{figure}

\normalsize{
We represent in Figure \ref{fig:exoplanet} the distribution of detected bound exoplanets and very low mass companions in the plane of the host and companion masses .
The two solutions for this event are shown by purple circles.
The planets detected by microlensing are represented as red circles, in which open and filled circles indicate planets whose masses were estimated by Bayesian analyses and directly measured by higher order microlensing effects, respectively.
The green circles indicate the microlensing binary events with large mass ratios of $q>0.1$, which are brown-dwarf binaries or brown dwarfs orbiting around VLMS (\citealp{2013ApJ...768..129C}; \citealp{2015ApJ...798..123J}; \citealp{2016ApJ...822...75H}; \citealp{2017ApJ...843...59H}).  
They are likely formed by a different mechanism from that of planetary systems with $q<0.03$.
The binary model of this work belongs to this group.
The favored planetary models can be one of a new class of planetary system, having an extremely low mass host with a planetary mass ratio ($q<0.03$).
A similar system is MOA-2011-BLG-262L, where one of the models has a sub-Earth mass moon orbiting a gas giant primary (\citealp{2014ApJ...785..155B}).
And, \citet{2017Natur.548..183M} also found several short-timescale binary events which could have very low mass hosts.
If such low host mass planetary systems exist, this offers a challenge to planet formation theory.
However, the priors in our Bayesian analysis are highly uncertain at low masses for two reasons.
First, the mass function in the low-mass region is highly uncertain, and second, we have no idea if the low mass objects are likely to host planetary mass ratio companions.
Thus, we need to measure the lens mass and distance to know the abundances of planets in these low mass primaries. 
\vspace{0.2in}

In order to improve the situation to measure the physical parameters of such short-timescale microlensing events, firstly we have to get better data coverage and accuracy in the light curves to distinguish between competing models.
Secondly,  we can constrain the lens mass from the lens flux measurements through high resolution Adaptive Optics (AO) imaging from the ground or the HST directly. 
Although we will not be able to detect the flux from such a low mass object, we will be able to find upper limit constraints on the lens mass. 
Thirdly, the mass can be measured by observing microlensing parallax.
Because annual significant parallax can not be measured in these short time scale events, we should attempt to measure the terrestrial parallax effect (\citealp{2009ApJ...698L.147G}; \citealp{2009ApJ...703.2082Y}) as this would allow us to determine the lens mass directly from only the light curve when also combining finite source effects (\citealp{2015ApJ...799..181F}).  
The small Einstein radius of the low mass object is suitable for observing this effect given the short baseline of observatories spread across the Earth's surface.  
This would be implemented more frequently through improvements of microlensing surveys (\citealp{2012SPIE.8444E..47P}) for rapid event identification and follow-up organization (\citealp{2013PASP..125.1031B}).  
Space-based parallax is more powerful in order to measure the lens mass and this has been demonstrated by the simultaneous observations by the Spitzer satellite and ground-based telescopes (\citealp{2014ApJ...784...64G}; \citealp{2016ApJ...819...93S}). 
\vspace{0.2in}

In addition, we also advocate for the importance of considering all viable solutions in all ambiguous events which can be interpreted by both a planetary and a stellar binary lens system (\citealp{2004ApJ...611..528G}; \citealp{2012ApJ...756...48C}; \citealp{2014ApJ...787...71P}; \citealp{2018arXiv180210246H}) in order to conduct comprehensive microlensing statistical analyses, such as that of \citet{2016ApJ...833..145S} and \citet{2018arXiv180202582U}.
Missing the solutions could lead to incorrect statistical results so that we should fully examine all the solutions in all microlensing events including ambiguous events, particularly in short-timescale events.

\vspace{0.2in}

NASA's WFIRST mission (\citealp{2012arXiv1208.4012G}; \citealp{2013arXiv1305.5422S}) will conduct a space-based microlensing survey toward the Galactic bulge in the near infrared and is planned to launch in mid 2020s.  
WFIRST fulfills all of these requirements with its high precision, high cadence and high spatial resolution survey. 
It is very sensitive to such short-timescale planetary events and will enable us to determine the lens masses directly using the space parallax.
The PRIME (PRime-focus Infrared Microlensing Experiment) project is planning to conduct a microlensing survey toward the central region of the Galactic bulge ($|b| \le 2 {\rm deg}$) in {\it H}-band by using a new 1.8m wide FOV telescope in South Africa. 
As the PRIME telescope can observe the same fields as WFIRST, simultaneously, the space-based parallax will be observed regularly without the current alert and follow-up strategy. 
The longer coverage of the PRIME survey can also add time series outside of the WFIRST's observing window of 72 days duration.
These future microlensing surveys will reveal the frequency of exoplanets around low-mass brown dwarfs and even the frequency of planetary mass objects orbited by exo-moons.
}

\section{Summary And Conclusions}\label{sec:SAC}
\normalsize{
We analyzed the short-timescale microlensing event MOA-2015-BLG-337 and found there exist two very degenerate solutions with mass ratios of $q\sim10^{-2}$ and $q\sim10^{-1}$.
The former planetary solutions are favored over the latter binary solution by $\Delta\chi^{2}\sim6$.
There are degeneracies between the close and the wide solutions with $\Delta\chi^{2}\simeq0.2$ and $\Delta\chi^{2}\simeq2.0$ in the planetary and the binary models, respectively.
We could measure the finite source effect for only the planetary models and obtained a large value of $\rho\sim10^{-2}$ implying a very low mass lens.
We could not detect a significant microlens parallax signal in the light curve. 
Therefore, we conducted a Bayesian analysis by applying the observed $t_{\rm E}$ and $\theta_{\rm E}$ values to estimate the probability distribution of the physical properties of the lens system.
Since the host is likely to have a lower mass than main sequence stars, we used a mass function extending to brown dwarf and planetary masses as a prior.
The Bayesian results for binary models do not depend greatly on the value of $\alpha_{\rm bd} $ and $\alpha_{\rm pl}$.
On the other hand, the results for planetary models strongly depend on $\alpha_{\rm pl}$, while the results are consistent with each other within $1\sigma$. 
So, there are two competing models of the lens system: (1) a brown dwarf/planetary mass boundary object orbited by a super-Neptune (the planetary model with $\alpha_{\rm pl}=0.49$) and (2) a brown dwarf binary (the binary model).
}

\acknowledgments
This research has made use of the KMTNet system operated by the Korea Astronomy and Space Science Institute (KASI) and the data were obtained at three host sites of CTIO in Chile, SAAO in South Africa, and SSO in Australia.
The OGLE project has received funding from the National Science Centre, Poland, grant MAESTRO 2014/14/A/ST9/00121 to AU.
Work by CR was supported by an appointment to the NASA Postdoctoral Program at the Goddard Space Flight Center, administered by USRA through a contract with NASA.
Work by N.K. is supported by JSPS KAKENHI Grant Number JP15J01676. 
Work by Y.H. is supported by JSPS KAKENHI Grant Number JP1702146.
NJR is a Royal Society of New Zealand Rutherford Discovery Fellow. 
This work was supported by JSPS KAKENHI Grant Number JP17H02871.
Work by C.H. was supported by the grant (2017R1A4A1015178) of National Research Foundation of Korea.
Work by WZ, YKJ, and AG were supported by AST1516842 from the US NSF. WZ, IGS, and AG were supported by JPL grant 1500811.
{} 

\begin{deluxetable*}{cccccc}[htb!]
\tablecaption{Data Sets for MOA-2015-BLG-337}
\tablecolumns{5}
\tablenum{1}
\tablewidth{0pt}
\tablehead{
\colhead{Telescope} &
\colhead{Diameter} &
\colhead{band} & \colhead{Number of data} & \colhead{$k$\tablenotemark{1}}  \\
\colhead{} & \colhead{(m)} &
\colhead{} & \colhead{} & \colhead{} 
}
\startdata
MOA-II & 1.8 & $R+I$ & 13813 & 1.081 \\ 
&  & $V$ & 239 & 1.067 \\
OGLE-IV & 1.3 & $I$ & 1754 & 1.247 \\
B$\&$C  & 0.61 & $g$ & 124 & 0.898 \\
&  & $i$ & 146 & 1.569 \\
 & & $r$ & 145 & 1.009 \\
KMTNet-SAAO & 1.6 & $I$ & 1555 & 1.576 \\
KMTNet-SSO & 1.6 & $I$ & 1194 & 1.702 \\
KMTNet-CTIO & 1.6 & $I$ & 2846 & 1.589 \\ 
\enddata
\label{table:data}
\tablenotetext{1}{
The coefficients for error renormalization, see text.
}

\end{deluxetable*}

\begin{deluxetable*}{c c c c c c c }[htb!]
 \tablecaption{The Best-fit Model Parameters\label{tab:modelparameters}}
 \tablecolumns{7}
 \tablenum{2}
 \tablewidth{0pt}
\tablehead{
\colhead{Parameters} & \colhead{Units} & \colhead{Planetary} & \colhead{Planetary} & \colhead{Binary} & \colhead{Binary} & \colhead{BSPL}\\
\colhead{} & \colhead{} & \colhead{(close)} & \colhead{(wide)} & \colhead{(close)} &  \colhead{(wide) } 
}
\startdata
$t_{\rm E}$ & days & 5.382(94) & 5.432(90) & 5.496(95) & 6.104(99) & 6.052 \\
$t_{0}$ &  HJD-2450000 & 7214.9183(20) &7214.9243(23) &  7214.9120(21) & 7215.7390(21) & 7215.0536\\
$u_{0}$ &  & 0.0470(11) & 0.0567(18) & 0.0458(12) & 0.0435(13) & 0.0331\\
$q$ & & 0.0108(14) & 0.0109(12) & 0.178(23) & 0.235(54) & -\\
$s$ & & 0.606(33) & 1.548(64) & 0.263(11) & 4.71(38) & -\\
$\alpha$ & radian & 1.4445(57) & 1.4449(63) & 2.5359(78) & 0.5877(75) & -\\
$\rho$ & & 0.0223(30) & 0.0263(23) & 0.0166($<0.0183$)\tablenotemark{a} & 0.0152$(<0.0176)\tablenotemark{a}$ & 0.0041\\ 
$t_{0,2}\tablenotemark{b}$ & HJD-2450000 & - & - & - & - & 7214.7049\\
$u_{0,2}\tablenotemark{b}$ & & - & - & - & - & 0.0379\\
$\rho_{2}\tablenotemark{b}$ & & - & - & - & - & 0.0048\\ \hline\hline
fit $\chi^{2}$ & & 21800.18 & 21800.34 & 21806.26 & 21808.21 & 21852.51\\ 
\enddata
\tablecomments{The numbers in parentheses indicate the 1$\sigma$ uncertainties derived from the 16th/84th percentile values of the stationary distributions given by MCMC.}
\tablenotetext{a}{
This value indicates a $1\sigma$ upper limit on $\rho$. 
The binary model including the parameter $\rho$ is favored over one without $\rho$ by only $\Delta\chi^{2} <4$. 
This indicates that the finite source effects for the binary models are not significant for fitting, and it's questionable to accept this $\rho$ value directly. 
Thus, we do not adopt the best-fit $\rho$ value but put an upper limit on $\rho$.
}
\tablenotetext{b}{
These are secondary source parameters for binary-source point-lens (BSPL) model.
}
\label{model-parameter}
\end{deluxetable*}
 
\begin{deluxetable}{l c c c }[htb!]
\label{tab:source}
\tablecaption{The Source Magnitude, Color and Angular Radius}
\tabletypesize{\footnotesize}
\tablecolumns{4}
\tablenum{3}
\tablewidth{0pt}
\tablehead{
 & \multicolumn{1}{c}{$I$ (mag)} & \multicolumn{1}{c}{$V-I$ (mag)} &\multicolumn{1}{c}{$\theta_{*}$ ($\mu$as)}
}
\startdata
Source (measured from the light curve) & $19.682\pm0.012$ & $1.611\pm0.019$ \\
Source\tablenotemark{1} (intrinsic) & $18.478\pm0.064$ & $0.621\pm0.074$ & $0.621\pm0.045$ \\
Source\tablenotemark{2} (intrinsic) & $18.598\pm0.064$ & $0.720\pm0.074$ & $0.606\pm0.045$ \\
\enddata
\tablenotetext{1}{This value is obtained by assuming the source suffered from the same extinction and reddening for the RCG.}
\tablenotetext{2}{This value is obtained by assuming the source suffered from the extinction and reddening of 0.9 times as much as that for the RCG.}
\end{deluxetable}

\begin{deluxetable}{c | c c c | c c c }[htb!]
\tablecaption{Lens Properties derived from the Bayesian analysis with $\alpha_{\rm bd}=0.49$}
\tabletypesize{\footnotesize}
\tablecolumns{5}
\tablenum{4}
\tablewidth{0pt}
\tablehead{
\hline
\multicolumn{1}{c|}{Lens Parameters} & \multicolumn{3}{c|}{$\alpha_{\rm pl}=-4.0$} & \multicolumn{3}{c}{$\alpha_{\rm pl}=0.49$} \\ 
& \multicolumn{1}{c}{Planetary} & \multicolumn{2}{c|}{Binary}& \multicolumn{1}{c}{Planetary} & \multicolumn{2}{c}{Binary} \\
& \multicolumn{1}{c}{} & \multicolumn{1}{c}{$(s<1)$} & \multicolumn{1}{c|}{$(s>1)$} & \multicolumn{1}{c}{} & \multicolumn{1}{c}{$(s<1)$} & \multicolumn{1}{c}{$(s>1)$}
}
\startdata
$M_{\rm host}$ & $25.4_{-14.2}^{+57.5} M_{\rm{Jup}}$ & \multicolumn{2}{c|}{$75.5_{-45.3}^{+126.6} M_{\rm{Jup}}$} & $9.8^{+37.6}_{-6.8} M_{\rm{Jup}}$ & \multicolumn{2}{c}{$76.4_{-45.1}^{+127.9} M_{\rm{Jup}}$}\\
$M_{\rm comp}$ & $87.2_{-48.6}^{197.4} M_{\oplus}$ & \multicolumn{2}{c|}{$17.8_{-10.7}^{+29.8} M_{\rm{Jup}}$} & $33.7^{+129.0}_{-23.2} M_{\oplus}$ & \multicolumn{2}{c}{$18.0_{-10.6}^{+30.1} M_{\rm{Jup}}$}\\
$r_{\perp}$ & $0.28^{+0.06}_{-0.10}$ AU & $0.19_{-0.06}^{+0.07}$ AU & $3.2_{-1.0}^{+1.2}$ AU &$0.24^{+0.06}_{-0.09}$ AU & $0.19_{-0.06}^{+0.07}$ AU & $3.3_{-1.0}^{+1.2}$ AU \\
$D_{\rm L}$ & $7.4^{+1.1}_{-1.0}$ kpc & \multicolumn{2}{c|}{$6.3_{-1.3}^{+1.2}$ kpc} & $  7.1^{+1.1}_{-1.0}$ kpc & \multicolumn{2}{c}{$6.3_{-1.3}^{+1.2}$ kpc} \\
\enddata
\tablecomments{
These results are obtained by equation (\ref{eq:power-law}) with $\alpha_{\rm bd}=0.49$ and $M_{\rm break}=0.7$ as a prior initial mass function (IMF).
}
 \label{tab:Sumi-IMF}
\end{deluxetable}

\begin{deluxetable}{c | c c c | c c c}[htb!]
\tablecaption{Lens Properties derived from the Bayesian analysis with $\alpha_{\rm bd}=0.8$}
\tabletypesize{\footnotesize}
\tablecolumns{5}
\tablenum{5}
\tablewidth{0pt}
\tablehead{
\hline
\multicolumn{1}{c|}{Lens Parameters} & \multicolumn{3}{c|}{$\alpha_{\rm pl}=-4.0$} & \multicolumn{3}{c}{$\alpha_{\rm pl}=0.8$} \\ 
& \multicolumn{1}{c}{Planetary} & \multicolumn{2}{c|}{Binary}& \multicolumn{1}{c}{Planetary} & \multicolumn{2}{c}{Binary} \\
& \multicolumn{1}{c}{} & \multicolumn{1}{c}{$(s<1)$} & \multicolumn{1}{c|}{$(s>1)$} & \multicolumn{1}{c}{} & \multicolumn{1}{c}{$(s<1)$} & \multicolumn{1}{c}{$(s>1)$}
}
\startdata
$M_{\rm host}$ & $19.7^{+43.2}_{-9.3} M_{\rm{Jup}}$ & \multicolumn{2}{c|}{$64.3_{-39.9}^{+114.0} M_{\rm{Jup}}$} & $6.3^{+19.6}_{-3.9}  M_{\rm Jup}$ & \multicolumn{2}{c}{$62.0_{-40.4}^{+113.5} M_{\rm Jup}$} \\
$M_{\rm comp}$ & $67.7^{+148.5}_{-31.9} M_{\oplus}$ & \multicolumn{2}{c|}{$15.1_{-9.4}^{+26.8} M_{\rm{Jup}}$} & $21.6^{+67.3}_{-13.4} M_{\oplus}$ & \multicolumn{2}{c}{$14.6_{-9.5}^{+26.7} M_{\rm Jup}$}\\
$r_{\perp}$ & $0.28^{+0.06}_{-0.10}$ AU & $0.17_{-0.06}^{+0.07}$ AU & $3.1_{-1.0}^{+1.2}$ AU & $0.23^{-0.08}_{+0.06}$ AU & $0.17_{-0.06}^{+0.07}$ AU & $3.1_{-1.1}^{+1.3}$ AU\\
$D_{\rm L}$ & $7.4^{+1.0}_{-1.0}$ kpc & \multicolumn{2}{c|}{$6.2_{-1.3}^{+1.2}$ kpc} & $7.0^{+1.1}_{-1.0}$ kpc & \multicolumn{2}{c}{$6.2_{-1.4}^{+1.2}$ kpc}  \\
\enddata
\tablecomments{
These results are obtained by equation (\ref{eq:power-law}) with $\alpha_{\rm bd}=0.8$ and $M_{\rm break}=0.5$ as a prior IMF.
}
 \label{tab:Mroz-IMF}
\end{deluxetable}

\end{document}